\documentstyle{article}
\input epsf
\title{Numerical Study of Competing Spin-Glass and Ferromagnetic Order}
\author{M.V. Simkin \\ {\em Department of Physics, Brown University,}
\\{\em Providence, RI 02912-1843}}
\date{ }
\begin{document}
\maketitle
\begin{abstract}
Two and three dimensional random Ising  models with a Gaussian distribution of
couplings with variance $J$ and non-vanishing mean value $J_0$ are studied 
using the zero-temperature domain-wall renormalization group (DWRG). 
The DWRG trajectories in 
the ($J_0,J$) plane after rescaling  can be collapsed on two curves: one for 
$J_0/J > r_c$ and other for $J_0/J < r_c$. In the first case the DWRG flows
are toward the ferromagnetic fixed point both in two and three dimensions 
while in the second case flows are towards a paramagnetic fixed point and 
spin-glass fixed point in two and three dimensions respectively. No evidence
for an extra phase is found.
\end{abstract}

In some range of concentration of magnetic impurities in a non-magnetic
host one observes a competition between spin-glass and ferromagnetic order
\cite{fh}. The phenomenon can be described by an Ising
model in which couplings are distributed randomly with some non-vanishing
mean value $J_0$ and width $J$. This issue has been addressed  by 
Sherrington and Kirkpatrick  \cite{sk} in the case of a soluble 
infinite-range model.  For the case of insoluble short range model 
Migdal-Kadanoff real-space rescaling \cite{ys}, \cite{sy} and computer 
simulations \cite{k}-\cite{uo} have been employed.  
It is established \cite{mc1},\cite{bm} that there is no 
finite-temperature phase transition in a two-dimensional (2d) Ising spin-glass 
(random Ising model with $J_0=0$). McMillan  \cite{mc} investigated
the 2d random Ising model for the general case, $J_0$ not necessarily zero,    
using finite-temperature domain-wall renormalization group (DWRG).
He found  at small temperatures only two phases: ferromagnetic and 
paramagnetic. However, some later studies 
\cite{oz}, \cite{uo} have found a finite temperature transition
to a ``random antiphase state'', which has similar properties to a spin glass.
Therefore more study of the subject is desirable.
To my knowledge no DWRG study of three dimensional (3d) random Ising model 
(with $J_0 \ne 0$) have been done.
In this paper the 2d and 3d random Ising models are investigated 
using the zero-temperature DWRG \cite{mc}-\cite{bm}. In 
particular no evidence for the ``random antiphase state'' is found.

The system is the Edwars-Anderson model \cite{ea} of an Ising spin glass 
with Hamiltonian:
\begin{equation}
H=-\sum_{ij}J_{ij}S_iS_j,
\end{equation}
where $S_i= \pm 1$, $J_{ij} \ne 0$ only when lattice sites $i,j$ are nearest 
neighbors. Each coupling constant $J_{ij}$ is an independent Gaussian 
distributed 
random variable with $P(J_{ij})=\frac{1}{\sqrt{2 \pi J^2}}\exp(-(J_{ij}-J_0)^2/
2J^2)$. The  lattices studied are $d$-dimensional square ($d=2$) and simple 
cubic ($d=3$) lattices of linear size $L$.
The zero-temperature DWRG method \cite{mc1}-\cite{bm} was used by 
computing 
the ground-state energies for periodic and antiperiodic boundary conditions 
(BC) in one direction, with free BC in the other $d-1$ directions. The 
difference $\Delta E(L) \equiv E_p(L)-E_{ap}(L)$ 
is the domain-wall energy which may be  interpreted as an effective coupling
constant on scale $L$. $\Delta E(L) $ has sample to sample fluctuations
and one can define the mean $J_0(L)$ and the width $J(L)$ at scale $L$ by
\begin{eqnarray}
J_0(L)=<\Delta E(L)> \nonumber \\
J(L)=\sqrt{<(\Delta E(L)-J_0(L))^2>},
\end{eqnarray}
where $< \ldots >$ is the average over samples with different realizations
of the coupling constants $\{J_{ij}\}$.
The ground state energy is found by simulated annealing \cite{kgv}.
Each annealing was start from random spin configuration or $T = \infty$.
The temperature was reduced in $n_s$ steps from $T_h=5$ to $T_l=0.05$
by  a  factor $10^{-2/n_s}$. At each intermediate temperature $T_n =
T_h 10^{-2n/n_s}$ the spins are updated by a single trial flip using the 
Metropolis algorithm. 
As a single annealing is not guaranteed to reach the minimum energy state,
this procedure was repeated $n_a$ times, with the required CPU time 
proportional to $n_a \times n_s$. It is found empirically that the lowest 
energies for fixed CPU time are found when $n_a \simeq n_s$.
This annealing schedule is substantially better than a large number of
instantaneous quenches, for the same CPU time.
To take maximum advantage of vectorization  64  annealings were performed
simultaneously. 
To estimate the necessary values of $n_a$ and $n_s$  a few samples are tested
until increasing $n_a$ and $n_s$  no longer leads to lower energy. Even so
there is no guarantee that theses values of $n_a$ and $n_s$ are sufficient
for all of the samples. To estimate the errors due to this, two runs for the 
same set of one hundred samples with periodic BC but with different sequences 
of random numbers for the Metropolis spin updating are performed. The two 
energy minima $E_1,E_2$ found in the two nominally identical annealings were 
recorded. If true absolute minima are found than $E_1=E_2$ but, if the true 
minima are not reached, then $<(E_1-E_2)^2>$ is a measure of the error due to 
not finding the exact minimum. 
The numbers $n_a$ and $n_s$ are then increased until this error is less than
statistical error due to the finite number of samples, i.e. until $\delta E
/ E < N^{-1/2}$. All data is obtained for $N=10^4$ and the number of 
annealings $n_a$ and number of steps $n_s$ ranged from $n_a=64$, $n_s=10$ for
the smallest sizes (L=2) to $n_a=192$, $n_s=200$ for the largest ($L=9$ in 
$d=2$, $L=5$ in $d=3$).

The results of the simulations are shown in Fig. 1  as renormalization
group flows in the $(J_0,J)$ plane. All initial couplings $J_0(1)$ and
$J(1)$ are on the $J_0+J=1$ line from which the trajectories, 
indicated by 
dashed lines, start. Next symbols on the trajectories correspond to 
$L=2,3 \ldots$.

All the trajectories collapse into two curves by $L$ - independent rescaling 
\begin{eqnarray}
J_0(L) \rightarrow \lambda(r)J_0(L)  \nonumber \\
J(L) \rightarrow \lambda(r)J(L).
\end{eqnarray}

In Eq.3 the rescaling factor $\lambda(r)$ depends only on the ratio 
$r \equiv J_0/J$, and upon performing this rescaling all the flows
of Fig.1 collapse on to the flows of Fig.2. That the rescaling of Eq. 3 works
means that (at zero temperature) on any length scale $L$ the system is 
described by the Hamiltonian
of Eq. 1 and no other terms are generated by the RG transformation.

The flows for $r>r_c$ seem to
be flowing to a ferromagnetic fixed point with $J(L)/J_0(L)$ decreasing and
$J_0(L)$ increasing with increasing $L$ and one may speculate that the 
$L= \infty$ fixed point is at $J(\infty)/J_0(\infty)=0$,$J_0(\infty)=\infty$. 
For
$r<r_c$, on the other hand, the flows would seem to be different in $d=2$
and $d=3$. In $d=2$ the flow is shown in Fig.2(a) and it seems that as $L$
increases both $J_0(L)$ and $J(L)$ ultimately decrease to zero with 
$J(L)/J_0(L) \rightarrow \infty$. This would be interpreted as a paramagnetic
fixed point for any finite temperature and a spin-glass fixed point at
$T=0$. In $d=3$ (see Fig.2(b)) the couplings $J_0(L) \rightarrow 0$ and
$J(L) \rightarrow \infty$ which has the interpretation of a pure spin-glass
fixed point. This spin-glass at $T=0$ is believed to survive at low
temperatures $T < T_c$ in $d=3$ \cite{mcbm}.
Note that the critical value of $r=J_0/J$ is $r_c^{2d} \cong 1.05$ and 
$r_c^{3d} \cong 0.7$ which is in fairly good agreement with the values 
$r_c^{2d} \cong 1.2$ and $r_c^{3d} \cong 0.65$ obtained by a Migdal-Kadanoff
RG method \cite{sy}. The value of $r_c^{2d}$ is  in agreement with 
the result of McMillan \cite{mc}, who extrapolated his finite-temperature
DWRG data to $T=0$ and got $r_c^{2d} \cong 1.04$.

In a study of a related two-dimensional Ising model \cite{oz}, \cite{uo}
with a bimodal distribution $P(J_{ij})=p\delta(J_{ij}-1)+(1-p)\delta(J_{ij}+1)$
it was suggested that $J_0(L)$ and $J(L)$ scale as
\begin{eqnarray}
J_0(L) \sim L^a \nonumber \\
J(L) \sim L^{\tilde{a}},
\end{eqnarray}
where for some range of $p$, $a < 0 $ and $\tilde{a}>0$ (see \cite{oz},Fig. 9).
This was interpreted as evidence for a ``random antiphase'' state
with properties similar to those of spin-glass. The raw data of Fig.1(a)
seems to support the conjecture of Eq.4 as there are trajectories along which
$J_0(L)$ and $J(L)$ seem to scale in this way. However, the simple additional
rescaling of Eq.3 implies that the apparent $\tilde{a} > 0$  is a finite size 
effect which disappears at large $L$.

In conclusion this $T=0$ DWRG study of random Ising models with 
$<J_{ij}>=J_0 > 0$ indicates that, at low temperatures, there are only 
paramagnetic and ferromagnetic phases in $d=2$ and spin-glass and
ferromagnetic in $d=3$. No sign of any other phase is seen. 
Unfortunately, simulated annealing becomes ineffective for large system sizes
$L$, and computer limitations restricted $L$ to be quite small. However, 
recently it was found (Ref. \cite{bc}) that so-called branch and cut method 
allows 
one to study severall times larger systems than simulated annealing. 
Implementing this method for this particular problem may be a subject of 
future investigations.

I am grateful to  J.M. Kosterlitz for useful conversations and to 
M.J.P. Gingras for correspondence. This work was 
supported by National Science Foundation Grant No. DMR-9222812. 
Computations were performed on Cray EL98 at the Theoretical Physics Computing
Facility at Brown University.

\begin{figure*}
\epsfxsize= 10 cm \epsfbox{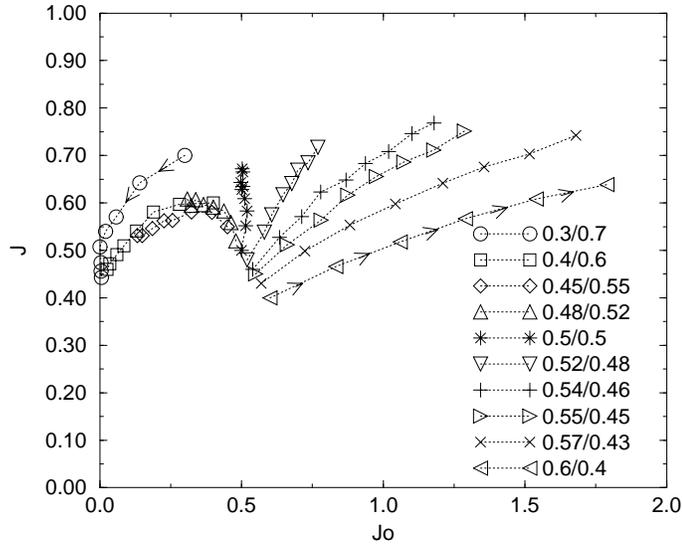}
\epsfxsize= 10 cm \epsfbox{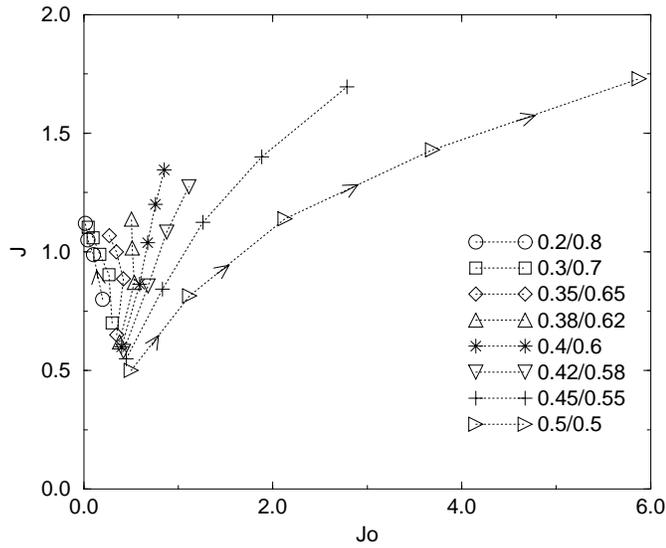}
\caption{Domain wall renormalization group trajectories for two (a) and three 
(b) dimensional
random Ising  models in the mean value ($J_0$) - variance ($J$) plane.  
The trajectories, indicated by dashed lines, start on the line $J_0+J=1$. 
Direction of flows is indicated by arrows on some of the trajectoriies.
The values of initial 
couplings $J_0$ and $J$ are indicated near corresponding symbols. Statistical
uncertainty on each point is about the size of the symbol.}
\end{figure*}
\begin{figure*}
\epsfxsize= 10 cm \epsfbox{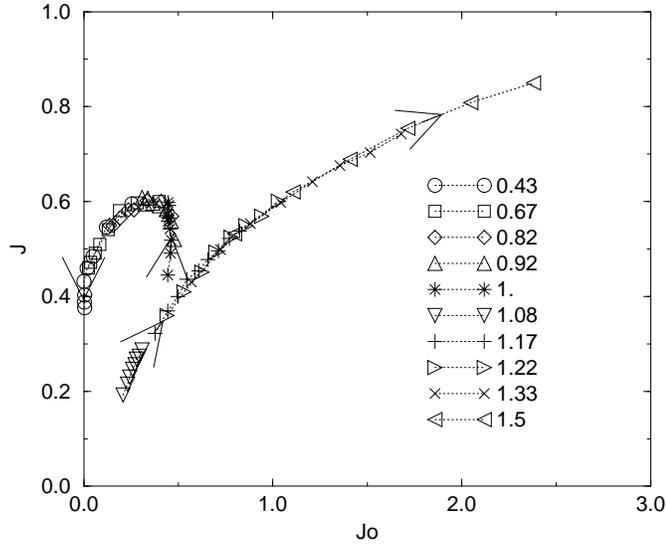}
\epsfxsize= 10 cm \epsfbox{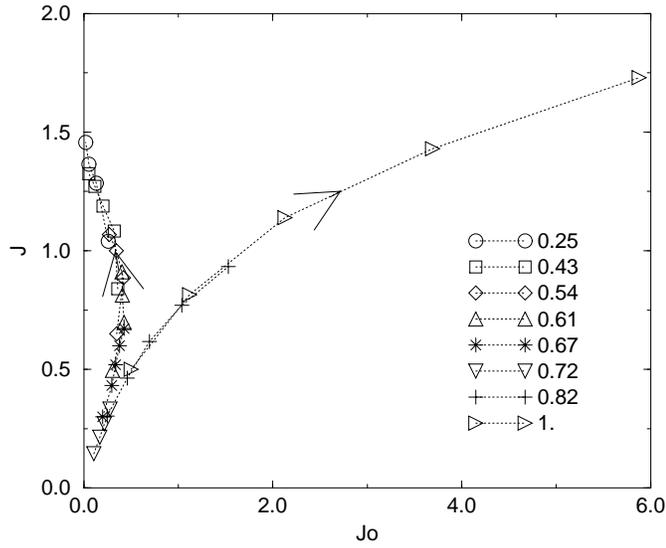}
\caption{Rescaled, using Eq.3, renormalization group trajectories of  
$d=2$ (a) and $d=3$(b)
random Ising  models. Instead of $J_0$ and $J$ as in Fig.1 their ratio
is indicated near corresponding symbols. Trajectories for $J_0/J > r_c$
($r_c^{2d} = 1.05$, $r_c^{3d} = 0.7$) flow to the ferromagnetic fixed
point, while those for $J_0/J < r_c$ flow to paramagnetic fixed point in
2d and spin-glass fixed point in 3d.}
\end{figure*}
\end{document}